# Net on-chip Brillouin gain
# based on suspended silicon nanowires

**Raphaël Van Laer⋆, Alexandre Bazin⋆, Bart Kuyken, Roel Baets and Dries Van Thourhout**

E-mail: `raphael.vanlaer@intec.ugent.be` or `alexandre.bazin@intec.ugent.be`

Photonics Research Group, Ghent University–imec, Belgium

Center for Nano- and Biophotonics, Ghent University, Belgium

⋆ These authors contributed equally to this work.

**Abstract.** The century-old study of photon-phonon coupling has seen a remarkable revival in the past decade. Driven by early observations of dynamical back-action, the field progressed to ground-state cooling and the counting of individual phonons. A recent branch investigates the potential of traveling-wave, optically broadband photon-phonon interaction in silicon circuits. Here, we report continuous-wave Brillouin gain exceeding the optical losses in a series of suspended silicon beams, a step towards selective on-chip amplifiers. We obtain efficiencies up to $10^4 \, \mathrm{W}^{-1}\mathrm{m}^{-1}$, the highest to date in the phononic gigahertz range. We also find indications that geometric disorder poses a significant challenge towards nanoscale phonon-based technologies.

## 1. Introduction

The interaction between photons and acoustic phonons has been investigated in bulk materials since the 1920s [1, 2]. In case the phonons are generated by optical forces, the interaction is often called stimulated Brillouin scattering (SBS) – a feedback loop in which energy flows from the optical waves to the mechanical oscillator. Its signature is narrowband amplification of an optical probe that is red-detuned by the phonon resonance frequency from a strong optical pump.

Although the mechanical linewidth does not exceed 100 MHz typically, there is no such inherent optical bandwidth restriction. Compared to cavity-based optomechanics [3–5], a circuit-oriented approach is intrinsically less power-efficient as the optical field is not resonantly enhanced. Nevertheless, the removal of the bandwidth restriction and the accompanying optical versatility has motivated a great deal of SBS work in small-core waveguides, from photonic crystal [6–8], dual-web [9] and subwavelength [10] fibres to chalcogenide [11–13] and silicon waveguides [14–16]. It may provide new integrated signal processing capabilities such as tunable RF notch filters [17] and true time delays [18]. The prospect is especially appealing in silicon photonic wires, whose strong confinement enhances the light-matter coupling. Mass-manufacturable silicon-on-insulator chips are therefore an exciting platform for high-density optomechanical circuitry, perhaps even at the quantum level [19–21].

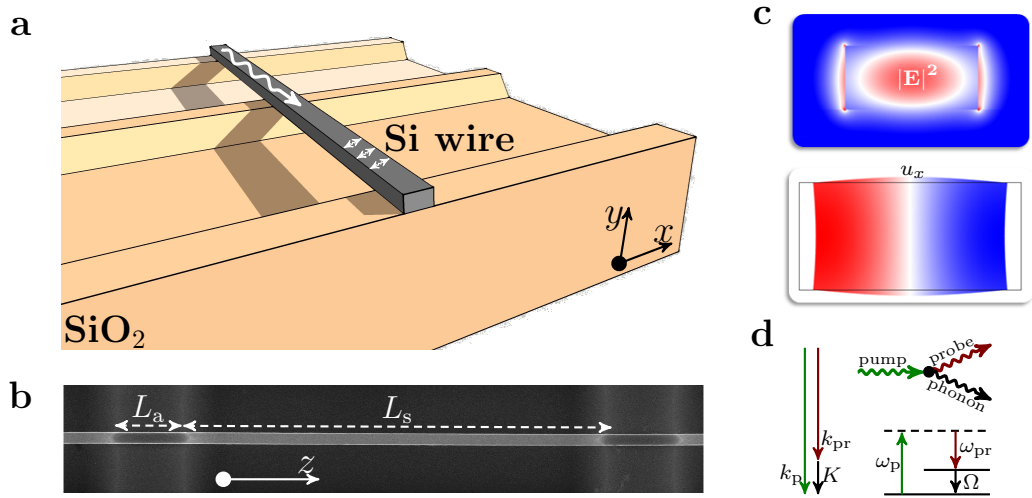

**Figure 1. A series of suspended silicon nanowires. a**, Impression of a silicon-on-insulator waveguide that consists of a series of suspensions and anchors. The photons propagate along the wire while the phonons are localized at their z-point of creation. **b**, Scanning electron micrograph of an actual suspension of length $L_s = 25.4\,\mu$m held by $L_a = 4.6\,\mu$m long anchors. **c**, Photonic (top) and phononic (bottom) traveling modes. **d**, The Brillouin process converts incoming pump photons with energy-momentum ($\hbar\omega_p, \hbar k_p$) into redshifted probe (Stokes) photons ($\hbar\omega_{pr}, \hbar k_{pr}$) and phonons ($\hbar\Omega, \hbar K$).

Recent work on this front has demonstrated promising photon-phonon coupling efficiencies in all-silicon waveguides [16]. The coupling was sufficiently strong to bring the waveguides into transparency, but phonon leakage and free-carrier absorption precluded actual amplification above the optical propagation loss. Here, we eliminate the phonon clamping loss – observing an increase of the phonon quality factor from 300 up to 1000 at room temperature – by fully suspending the silicon nanowires. Thus, we achieve a modest amount of gain exceeding the optical losses. The waveguides consist of a series of suspended beams, supported by silicon dioxide anchors (fig.1).

Finally, we observed a strong dependence of the phonon quality factor on the number of and distance between the suspensions. This indicates the presence of geometric disorder that broadens and splits the phonon dispersion relation in some cases, similar to Doppler broadening in gas lasers [22]. From a wider perspective and not limited to our system, such geometric disorder may hinder development of nanoscale phonon circuits quite generally [19–21, 23].

## 2. Results and discussion

### 2.1. Theoretical background

The following discussion is concerned with forward intra-modal scattering, in which co-propagating pump and probe waves generate low-wavevector, low-group-velocity acoustic phonons (fig.1d).

#### 2.1.1. Brillouin gain.
First, we briefly treat the small-signal Brillouin gain in a waveguide consisting of suspensions of length $L_s$ and anchors of length $L_a$. The section length is $L_{sec} = L_s + L_a$ and there are $N = \frac{L}{L_{sec}}$ such sections with $L$ the total waveguide length (fig.1). We denote the input pump power $P_p$ and the red-detuned probe (Stokes) power $P_{pr}$. As previously shown [16], the power $P_{pr}$ of the probe obeys

$$\frac{\mathrm{d}P_{pr}}{\mathrm{d}z} = -\left(\tilde{\mathcal{G}} P_p e^{-\alpha z}\Im\mathcal{L} + \alpha\right)P_{pr} \qquad\text{suspensions} \qquad (1)$$

$$\frac{\mathrm{d}P_{pr}}{\mathrm{d}z} = -\alpha P_{pr} \qquad\qquad\qquad\text{anchors}$$

in the low-cascading regime and with $\tilde{\mathcal{G}}$ the Brillouin gain coefficient, $\mathcal{L}(\Delta) = \frac{1}{-2\Delta+i}$ the complex Lorentzian, $\Delta = \frac{\Omega-\Omega_m}{\Gamma_m}$ the normalized detuning, $\Gamma_m$ the phonon linewidth and $\Im\mathcal{L} = -\frac{1}{4\Delta^2+1}$. To derive (1), we assumed that the phonon propagation loss far exceeds the photon propagation loss and that the photon-phonon coupling is weak relative to the spatial phonon decay [24]. In particular, in this work the photon decay length $\alpha^{-1}$ is about a centimeter, while the phonons spatially decay over a couple of nanometers in the $z$-direction [16]. Indeed, the flat dispersion of these Raman-like [7, 25] phonons yields an exceedingly low group velocity [16]. Therefore, each suspension consists of a series of independent mechanical oscillators – whose frequency depends on the local width [16]. This phonon locality also implies that the anchor does not contribute to the Brillouin gain.

We treat the optical loss $\alpha$ as distributed, although it may in fact be partially localized at the interfaces or be unequal in the suspensions and anchors. This is a good approximation as the SBS strength in the next section only depends on the remaining pump power, not on how some of it was lost in the previous sections. The ansatz $P_{pr} = g(z)e^{-\alpha z}$ and piecewise integration of (1) results in

$$\ln\frac{g(L)}{g(0)} = -\tilde{\mathcal{G}} P_p L_{s,eff}\Im\mathcal{L}\sum_{k=0}^{N-1} e^{-\alpha k L_{sec}} = -\tilde{\mathcal{G}} P_p L_{s,eff}\Im\mathcal{L}\frac{1-e^{-\alpha L}}{1-e^{-\alpha L_{sec}}} \approx -\tilde{\mathcal{G}} P_p f_s L_{eff}\Im\mathcal{L}$$

with the effective suspension length $L_{s,eff} = \frac{1-e^{-\alpha L_s}}{\alpha} \approx L_s$ since the sections are much smaller than the optical decay length ($L_{sec} \ll \alpha^{-1}$) and $f_s = \frac{L_s}{L_{sec}}$. Therefore we obtain

$$\ln\frac{P_{pr}(L)}{P_{pr}(0)e^{-\alpha L}} = \frac{\tilde{\mathcal{G}} P_p f_s L_{eff}}{4\Delta^2+1} \qquad (2)$$

These experiments are the circuit analog of cavity-based optomechanically induced transparency [26, 27]. However, our system features spatially stronger mechanical than optical damping, such that it is the optical response that is modified here [24].

*2.1.2. Cross-phase modulation.* Gain measurements provide access to all relevant optomechanical parameters, but require careful calibration of the on-chip pump power $P_p$. In contrast, a cross-phase modulation (XPM) measurement [14, 16] is, in absence of free-carriers, intrinsically calibrated: it provides access to the ratio of the photon-phonon coupling and the electronic Kerr effect independent of pump power. These experiments are the circuit analog of cavity-based coherent wavelength conversion [28], although the conversion need not take place between two optical resonances in our case.

We assume weak XPM and denote the envelopes of the pump and its red- and blue-detuned sidebands $a_p$, $a_{p-}$ and $a_{p+}$ and similarly for the injected probe $a_{pr}$ and the XPM-imprinted blue-shifted sideband $a_{pr+}$. The imprinted sideband grows as [16]

$$\frac{da_{pr+}}{dz} = -\frac{i}{2}\left(4\gamma_{Ks} + \tilde{\mathcal{G}}\mathcal{L}\right)\left(a_p a_{p-}^\star + a_{p+}a_p^\star\right)a_{pr} - \frac{\alpha}{2}a_{pr+} \qquad \text{suspensions}$$

$$\frac{da_{pr+}}{dz} = -i2\gamma_{Ka}\left(a_p a_{p-}^\star + a_{p+}a_p^\star\right)a_{pr} - \frac{\alpha}{2}a_{pr+} \qquad \text{anchors}$$

with $\gamma_K$ the Kerr parameter. Note that the XPM can also be seen as a four-wave mixing process with photon creations $(a^\star)$ and annihilations $(a)$ given by $a_{pr+}^\star\left(a_p a_{p-}^\star + a_{p+}a_p^\star\right)a_{pr}$. We assume that the pump and probe remain undepleted by the XPM, but include their absorptive decay. Then we get

$$\frac{da_{pr+}}{dz} = C_s e^{-\frac{3\alpha}{2}z} - \frac{\alpha}{2}a_{pr+} \qquad \text{suspensions}$$

$$\frac{da_{pr+}}{dz} = C_a e^{-\frac{3\alpha}{2}z} - \frac{\alpha}{2}a_{pr+} \qquad \text{anchors}$$

with $C_s = \frac{1}{2}\left(4\gamma_{Ks} + \tilde{\mathcal{G}}\mathcal{L}\right)C$, $C_a = 2\gamma_{Ka}C$ and $C = -i\left(a_p a_{p-}^\star + a_{p+}a_p^\star\right)a_{pr}|_{z=0}$. Inserting the ansatz $a_{pr+}(z) = g(z)e^{-\frac{\alpha}{2}z}$ and piecewise integrating $(g(0) = 0)$ yields

$$g(L) \approx (C_s L_s + C_a L_a)\sum_{k=0}^{N-1}e^{-\alpha k L_{sec}} = (C_s L_s + C_a L_a)\frac{1 - e^{-\alpha L}}{1 - e^{-\alpha L_{sec}}} \approx (C_s f_s + C_a f_a)\,L_{eff}$$

where we used $L_{a,eff} \approx L_a$, $L_{eff} = \frac{1 - e^{-\alpha L}}{\alpha}$ and $f_a = 1 - f_s$. Therefore,

$$|a_{pr+}(L)|^2 = 4\overline{\gamma}_K^2 \mathcal{F}|C|^2 L_{tot,eff}^2 e^{-\alpha L} \propto \mathcal{F}$$

with the averaged Kerr parameter $\overline{\gamma}_K = \gamma_{Ks}f_s + \gamma_{Ka}f_a$, the normalized Fano function

$$\mathcal{F}(\Delta) = \left|1 + f_s\frac{\tilde{\mathcal{G}}}{4\overline{\gamma}_K}\mathcal{L}(\Delta)\right|^2 = |1 + r\mathcal{L}(\Delta)|^2 \qquad (3)$$

and $r = f_s\frac{\tilde{\mathcal{G}}}{4\overline{\gamma}_K}$ the ratio between the mechanically- and Kerr-driven XPM.

Including two-photon absorption, free-carrier index changes and free-carrier absorption, (3) no longer holds. For instance, in case of two-photon absorption the Kerr parameter should be replaced by $\overline{\gamma}_{\mathrm{K}} \rightarrow \overline{\gamma}_{\mathrm{K}} - i\overline{\gamma}_{\mathrm{TPA}}$ and (3) becomes

$$\mathcal{F}(\Delta) = \left| 1 + e^{i\phi} f_s \frac{\tilde{\mathcal{G}}}{4\overline{\gamma}_{\mathrm{tot}}} \mathcal{L}(\Delta) \right|^2 = \left| 1 + e^{i\phi} r \mathcal{L}(\Delta) \right|^2 \tag{4}$$

with $\overline{\gamma}_{\mathrm{tot}} = \sqrt{\overline{\gamma}_{\mathrm{K}}^2 + \overline{\gamma}_{\mathrm{TPA}}^2}$ and $\tan\phi = \frac{\overline{\gamma}_{\mathrm{TPA}}}{\overline{\gamma}_{\mathrm{K}}}$. In our case, the two-photon $\phi$ is small and positive as $\frac{\overline{\gamma}_{\mathrm{TPA}}}{\overline{\gamma}_{\mathrm{K}}} \approx 0.1$. The free-carrier nonlinearity $\overline{\gamma}_{\mathrm{FC}}$, however, can give rise to negative $\phi$ (see Appendix).

## 2.2. Fabrication and passive characterization.

We started from air-cladded 220 nm thick, 450 nm wide silicon-on-insulator wires fabricated by 193 nm UV lithography (*www.ePIXfab.eu*) at imec. Next, we patterned an array of apertures in a resist spinned atop the wires. Then we immersed the chip in buffered hydrofluoric acid, which selectively etches the silicon dioxide substrate, until the wires were released. The end result was a series of suspended beams, each typically $25\,\mu\mathrm{m}$ long and held by $5\,\mu\mathrm{m}$ silicon dioxide anchors (fig.1). Simulations and measurements show that the reflections caused by these anchors are negligibly small (see Appendix). We found optical losses $\alpha \approx 5.5\,\mathrm{dB/cm}$ by the cut-back method, which are a factor 2 larger than before the etch. This is likely related to a deterioration of the wires' surface state and consistent with both (1) the measured drop in free-carrier lifetime (see Appendix) and (2) the decrease in free-carrier absorption found in the gain experiment (fig.2b).

## 2.3. Optomechanical experiments

In this section, we discuss the guided-wave optomechanical characterization of a series of suspended silicon nanobeams. We used the experimental set-ups presented in [16]. Our device is characterized by the suspension length $L_{\mathrm{s}}$, the anchor length $L_{\mathrm{a}}$, the section length $L_{\mathrm{sec}} = L_{\mathrm{s}} + L_{\mathrm{a}}$, the number of suspensions $N$, the total length $L = NL_{\mathrm{sec}}$, the total effective length $L_{\mathrm{eff}} = \frac{1 - e^{-\alpha L}}{\alpha}$, the suspended fraction $f_{\mathrm{s}} = \frac{L_{\mathrm{s}}}{L_{\mathrm{sec}}}$ and the waveguide width $w$. Unless stated otherwise, these parameters have values $L_{\mathrm{s}} = 25.4\,\mu\mathrm{m}$, $L_{\mathrm{a}} = 4.6\,\mu\mathrm{m}$, $L_{\mathrm{sec}} = 30\,\mu\mathrm{m}$, $N = 85$, $L = 2535\,\mu\mathrm{m}$, $L_{\mathrm{eff}} = 2168\,\mu\mathrm{m}$, $f_{\mathrm{s}} = 0.85$ and $w = 450\,\mathrm{nm}$. In some cases, our waveguides have a non-suspended input/output section before/after the cascade of suspended nanobeams. We take this into account when calculating input pump powers (gain experiment) or suspended fractions (XPM experiment).

### 2.3.1. Brillouin gain.
First, we measured the amplitude response of our system. We injected a weak probe red-detuned from a strong pump and retrieved the probe power as a function of detuning (fig.2a). As before [16], we find gain resonances around 9.1 GHz. The on/off gain increases with pump power (fig.2b) and reaches the transparency point $\tilde{\mathcal{G}}P_\mathrm{p} = \alpha$ around $P_\mathrm{p} = 30$ mW. Beyond this pump power, the ouput exceeds the input probe photon flux. At the maximum pump power of 39 mW, we obtain guided-wave cooperativities [24] of $\tilde{\mathcal{C}} = \frac{(f_\mathrm{s}\tilde{\mathcal{G}})P_\mathrm{p}}{\alpha} = 1.7$. This modest net gain of 0.5 dB (fig.2b) is a step towards selective on-chip amplifiers that could be used for homodyne detection, in order to eliminate the requirement of a phase-stabilized local oscillator [29].

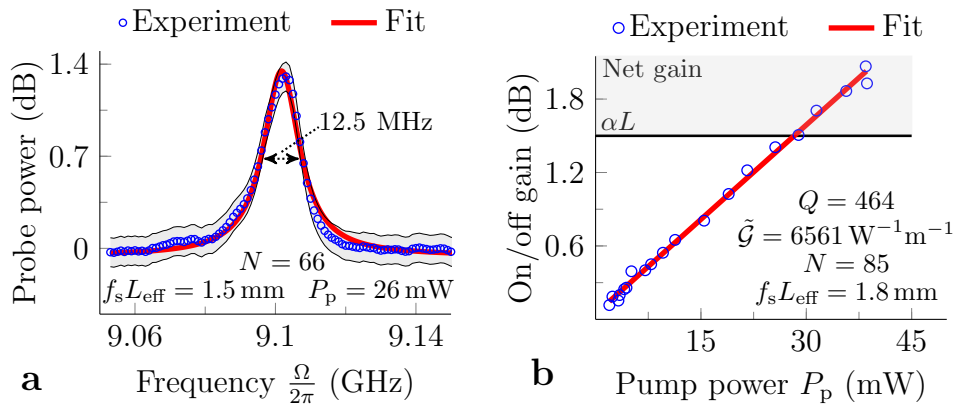

**Figure 2. Brillouin gain exceeding the optical losses. a**, An example of a Brillouin gain resonance, in this case with an on/off gain of 1.4 dB, quality factor of $Q_\mathrm{m} = 728$ and an on-chip input pump power of 26 mW. The shaded black area indicates uncertainty in the probe power. **b**, Scan of the on/off gain with pump power. At a pump power of 30 mW the transparency point is reached. For $P_\mathrm{p} > 30$ mW, more probe photons leave than enter the waveguide. The slope yields the Brillouin gain coefficient $\tilde{\mathcal{G}} = 6561\ \mathrm{W^{-1}m^{-1}}$ with a quality factor of 464 in this particular waveguide. Notably, the on/off gain scales linearly with pump power across the entire sweep – indicating the absence of free-carrier absorption in this range.

Notably, the linear scaling between on/off gain and pump power (fig.2b) indicates the absence of free-carrier absorption up to 40 mW [30,31]. In contrast, we previously measured increased nonlinear absorption already at 25 mW in silicon wires on a pillar [16]. Both this finding and the higher propagation losses (5.5 dB/cm instead of 2.6 dB/cm [16]) likely originate in a deterioration of the wires' surface state during the fabrication of the suspensions. In agreement with this hypothesis, we measured a drop in the free-carrier lifetime (see Appendix).

In case this structure were to be placed in a cavity, such as a silicon microring, it would also have to overcome coupling losses to achieve the photon/phonon lasing threshold ($\tilde{\mathcal{C}} > \mathcal{C} = 1$ [24]). We note that the acoustic linewidth ($\sim 10$ MHz) is a factor $10^2$ smaller than typical optical linewidths of silicon microcavities ($\sim 1$ GHz). Therefore, this would produce stimulated emission of phonons, not photons [4,5,24,32–34]. Such a device would not benefit from the spectral purification associated with Brillouin lasers [35,36]. The origin of this reversal of the damping hierarchy (going from waveguides to cavities) [24,34] lies in the exceedingly low group velocity of these Raman-like [7] acoustic phonons; indeed, despite enormously higher propagation losses they usually still have lower linewidths than photons [4,24,34]. Only uniquely high-quality optical

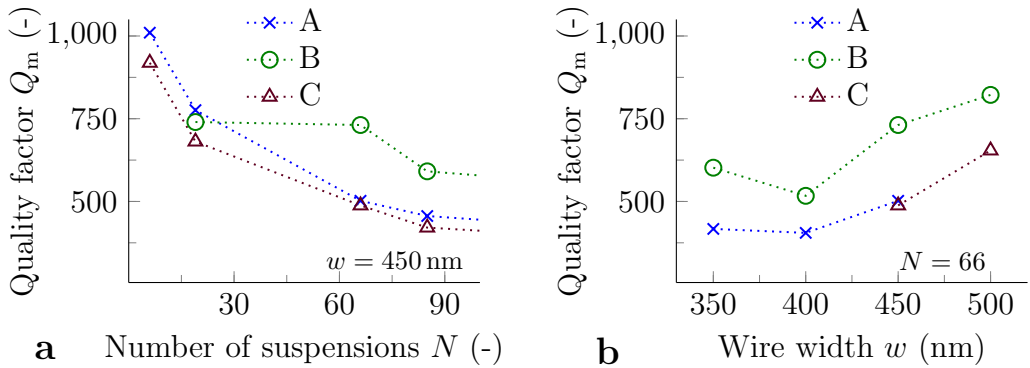

**a** Number of suspensions $N$ (-)   **b** Wire width $w$ (nm)

**Figure 3. The quality factor decreases with the number of suspensions.** We study the phonon quality factor for three samples (A, B and C) from the same wafer. The samples were designed to be identical. **a,** The quality factor increases up to 1010 when there are only 6 suspensions. For larger $N$, the quality factors approach $\approx 400$. Unless stated otherwise, all resonances are still well-fit by a Lorentzian function (fig.2a). **b,** In general, wider waveguides exhibit slightly larger $Q_{\mathrm{m}}$. However, this pattern is neither linear (samples A/C) nor monotonic (sample B). Some waveguides were defective, possibly because of a collapsed beam, and were excluded from the study.

cavities, to date realized only using silica [12, 37–40] or crystalline [41] materials, can produce lower photonic than phononic damping rates.

*2.3.2. Geometric disorder.* Next, we study the quality factor extracted from the gain resonances (fig.2a). We find that it strongly decreases with the number of suspensions $N$ (fig.3a); from $Q_{\mathrm{m}} \approx 10^3$ at $N = 6$ to $Q_{\mathrm{m}} \approx 400$ at $N = 85$. For larger numbers of suspensions in a spiral configuration ($N = 1332$, not shown), the quality factor levels off around $Q_{\mathrm{m}} \approx 340$. Notably, this relation changes from sample to sample – even if they originate from the same wafer (fig.3a). We attribute such variations to inhomogeneous broadening by geometric disorder, presumably in the width of the nanowires [16]. Indeed, the sensitivity of the resonance frequency $\frac{\Omega_{\mathrm{m}}}{2\pi}$ to width variations is $\frac{1}{2\pi}\frac{\mathrm{d}\Omega_{\mathrm{m}}}{\mathrm{d}w} \approx 20\,\mathrm{MHz/nm}$ [16]. Therefore, realistic width variations $\delta w$ of about 0.5 nm [42] yield inhomogeneous linewidths of about $\frac{\Gamma_{\mathrm{inh}}}{2\pi} \approx 10\,\mathrm{MHz}$ – comparable to those measured (fig.2a). Similar disorder has been studied in snowflake crystals [20].

Further, we investigated the influence of the width $w$ on the quality factor $Q_{\mathrm{m}}$. Since the resonance frequency scales inversely with width ($\frac{\Omega_{\mathrm{m}}}{2\pi} \propto w^{-1}$) [16], its sensitivity scales inverse quadratically with width ($\frac{\mathrm{d}\Omega_{\mathrm{m}}}{\mathrm{d}w} \propto w^{-2}$). Subsequently, the inhomogeneously broadened linewidth scales similarly ($\Gamma_{\mathrm{inh}} \propto w^{-2}$) in case the size of the width variations $\delta w$ does not depend on $w$. Then the quality factor scales linearly with width ($Q_{\mathrm{m}} = \frac{\Omega_{\mathrm{m}}}{\Gamma_{\mathrm{m}}} \propto w$). We indeed observe overall larger quality factors for wider wires, although this pattern is neither linear nor monotonic (fig.3b). We note that, unless stated otherwise, all resonances were still well-fit by a Lorentzian function (fig.2a). In case of sufficiently sampled geometric disorder, the gain curves would become convolutions of a Lorentzian and a probability function describing the geometric disorder (e.g. distribution of the width $w$). The largest deviations of such Voigt curves with respect to a Lorentzian occur in the tails (large $\Delta = \frac{\Omega - \Omega_m}{\Gamma_m}$), precisely where the relative uncertainty in the measured probe power is highest (fig.2a). Given this uncertainty, both Lorentzian and e.g. Gaussian-shaped curves produce good fits to the gain resonances. A low-temperature characterization would yield more information regarding the nature of the acoustic broadening mechanisms.

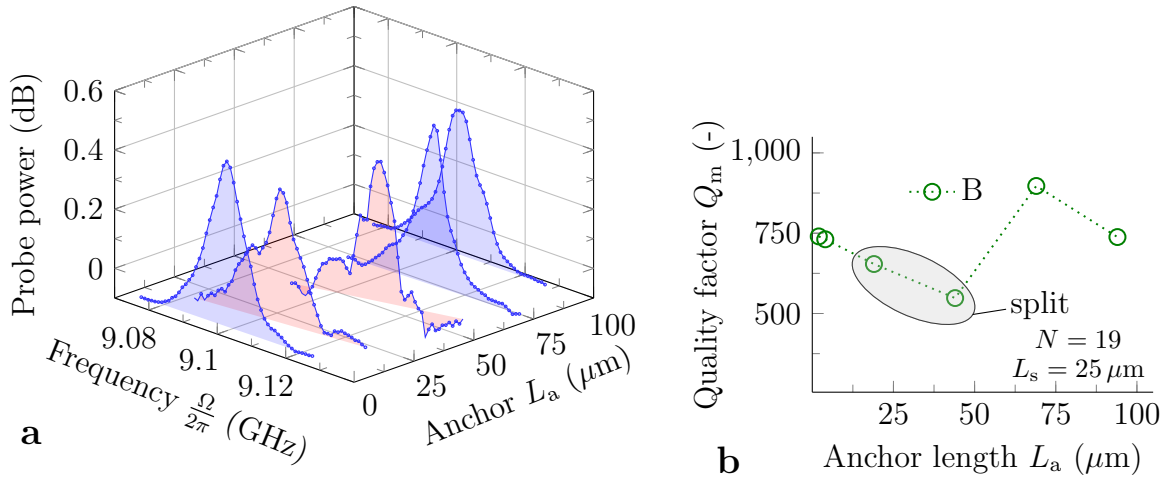

**Figure 4. The phonon resonance splits at certain anchor lengths. a**, As we sweep the anchor length, the initially clean curve splits at $L_a = 19\,\mu\text{m}$ and $44\,\mu\text{m}$ but recombines at $L_a = 69\,\mu\text{m}$. The pump power was $P_p = 26\,\text{mW}$ and the position of the first suspension was fixed in this sweep. **b**, A Lorentzian fit to the gain curves of (**a**) yields high quality factors at short and long anchors. We suspect that (**a**) and (**b**) arise from a nanometer-scale ($\delta w$) width fluctuation in this straight silicon wire.

There are two types of potential width fluctuations: (1) fast sidewall roughness with a coherence length $L_{coh}$ of only $\approx 50\,\text{nm}$ [43] and (2) slow variations in the average waveguide width $w$. We suspect that mechanism (2) is at play here, since even an individual section ($L_{sec} \approx 30\,\mu\text{m}$) is much larger than the coherence length ($\approx 50\,\text{nm}$ [43]) of the surface roughness. Therefore, sidewall roughness cannot explain the significant changes of $Q_m$ with $N$ (fig.3a): even a single section samples it fully ($L_{sec}/L_{coh} \approx 600$). In contrast, slow excursions of the waveguide width are consistent with such behavior. We confirm this by scanning the anchor length $L_a$ while keeping the number of suspensions $N$ and the suspension length $L_s$ constant (fig.4). In this sweep, the position of the first suspension is fixed. As $L_a$ increases, the initially clean resonance first splits at $L_a = 19\,\mu\text{m}$ and then recombines at $L_a = 69\,\mu\text{m}$ (fig.4a). Remarkably, the $L_a = 69\,\mu\text{m}$ wire even produces the highest quality factor (fig.4b). In light of the above discussion, this behavior likely stems from a nanometer-scale ($\delta w$) width excursion: short and long anchor waveguides avoid the width fluctuations and thus yield clean profiles. Both fig.3&4 are fingerprints of geometric disorder that hinders the development of integrated Brillouin-based technologies.

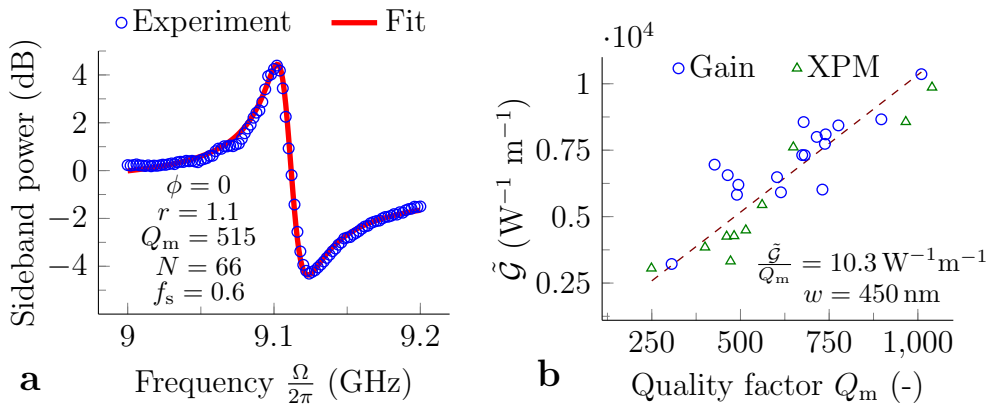

**Figure 5. The efficiency $\tilde{\mathcal{G}}$ reaches up to $10^4\,\mathrm{W^{-1}m^{-1}}$.** **a**, Example of a Fano resonance obtained from the XPM experiment, used to determine the quality factor $Q_{\mathrm{m}}$ and gain coefficient $\tilde{\mathcal{G}}$ given $\bar{\gamma}_{\mathrm{tot}} = 610\,\mathrm{W^{-1}m^{-1}}$ independently from the gain resonances. **b**, Plot of $(\tilde{\mathcal{G}}, Q_{\mathrm{m}})$-pairs for a large set of waveguides obtained from both the gain (fig.2a) and the XPM experiment (fig.5a). A linear fit without offset yields $\tilde{\mathcal{G}}/Q_{\mathrm{m}} = 10.3\,\mathrm{W^{-1}m^{-1}}$. Most variation results from uncertainty in the coupling efficiency ($\approx 25\%$). The two points at $Q_{\mathrm{m}} < 375$ concern a silicon wire on a pillar [16].

*2.3.3. Photon-phonon overlap.* Finally, we measure the XPM resonances (fig.5a) for a subset of waveguides to obtain an independent estimate of the photon-phonon interaction efficiency. Combined with the gain data, we obtain the $(\tilde{\mathcal{G}}, Q_{\mathrm{m}})$-pairs for a large set of waveguides (fig.5b) with fixed waveguide width $w = 450\,\mathrm{nm}$. A fit (fig.5b) to this dataset yields a non-resonant nonlinearity of $\frac{\tilde{\mathcal{G}}}{Q_{\mathrm{m}}} = 10.3\,\mathrm{W^{-1}m^{-1}}$ – in good agreement with earlier experiments [16] and predictions [25, 44]. The efficiencies reach up to $\tilde{\mathcal{G}} = 10360\,\mathrm{W^{-1}m^{-1}}$, the highest value obtained thus far in the gigahertz range.

## 3. Conclusion

Through a novel opto-acoustic nanodevice, a series of suspended silicon wires, we demonstrate modest (0.5 dB) net Brillouin gain with high efficiencies (up to $10^4\,\mathrm{W^{-1}m^{-1}}$). This device is a step towards integrated selective amplifiers. We find that fabrication disorder, likely in the waveguide width, broadens and splits the phonon resonances in some cases. In particular, the phonon quality factor strongly decreases as the number of suspended silicon beams increases. Such disorder is expected to hinder development of nanoscale phonon-based technologies quite generally – new techniques or better fabrication tools must be developed to address this issue.

## Appendices

*Influence of two-photon absorption and free-carriers on XPM*

In this section, we describe the influence of two-photon absorption (TPA) and free-carriers (FCs) on the Fano resonances. First, we recall that (see (3)), in absence of TPA

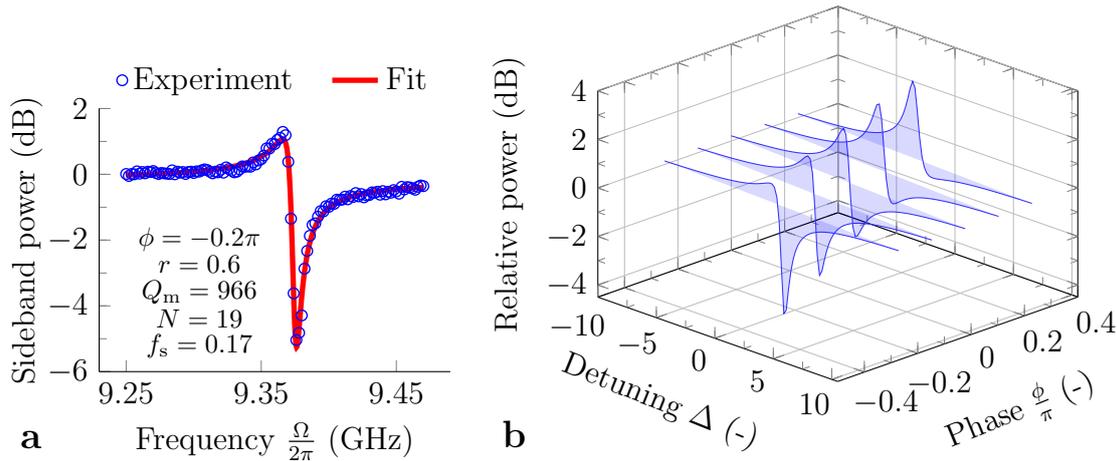

**Figure 6. Influence of phase $\phi$ on the Fano resonance. a,** In some cases, particularly for small $N$, we observe asymmetric ($\mathcal{F}_{\max}$ [dB] $\ll |\mathcal{F}_{\min}$ [dB]|) Fano resonances. The data is well-fit by including a phase shift $\phi < 0$ – physically linked to free-carrier generation (see (8)). **b,** Plot of the Fano function $\mathcal{F}(\Delta)$ as the phase shift $\phi$ is scanned ($r = 0.5$). The Fano resonance is symmetric ($\mathcal{F}_{\max}$ [dB] $= |\mathcal{F}_{\min}$ [dB]|) at $\phi = 0$; while for $\phi < 0$, the resonance becomes significantly deeper as in (**a**).

and FCs, the sideband power is proportional to $\mathcal{F}$ with

$$\mathcal{F}(\Delta) = \left| 1 + r \frac{1}{-2\Delta + i} \right|^2 \tag{5}$$

with $r = f_s \frac{\tilde{\mathcal{G}}}{4\tilde{\gamma}_K}$. One can show that this Fano function $\mathcal{F}$ has one maximum $\mathcal{F}_{\max}$ and one minimum $\mathcal{F}_{\min}$ given by

$$\mathcal{F}_{\max} = \frac{r^2 + 4 + r\sqrt{r^2 + 4}}{r^2 + 4 - r\sqrt{r^2 + 4}} = \frac{1}{\mathcal{F}_{\min}} \tag{6}$$

that are fully determined by $r$. This implies $\mathcal{F}_{\max}$ [dB] $= -\mathcal{F}_{\min}$ [dB], as evident in fig.5a. Inverting this for $r$ yields

$$r = \frac{\mathcal{F}_{\max} - 1}{\sqrt{\mathcal{F}_{\max}}} \tag{7}$$

Applied to fig.5a, we have $\mathcal{F}_{\max} = 4.4\,\mathrm{dB} = 2.75$ and thus $r = 1.1$ through (7) – in agreement with the $r$ obtained from a least-square fit (fig.5a). The extrema are reached at a detuning of $\Delta_{\max/\min} = \frac{1}{4}\left( r \mp \sqrt{r^2 + 4} \right)$. In the large $r$ limit, we get $\Delta_{\max} \to -\frac{1}{2r}$ and $\Delta_{\min} \to \frac{r}{2}$. In the small $r$ limit, we have $\Delta_{\max/\min} \to \frac{r \mp 2}{4}$. Therefore, the maximum XPM is always reached at a negative detuning between $-\frac{1}{2}$ and 0, typically ($r > 1$) close to the phonon resonance ($\Delta = 0$).

In some cases, we observe $\mathcal{F}_{\max}$ [dB] $\ll |\mathcal{F}_{\min}$ [dB]| – a clear indication that (5) is too simplistic. It turns out that the Fano function (5) must be replaced by [14]

$$\mathcal{F}(\Delta) = \left| 1 + e^{i\phi} r \frac{1}{-2\Delta + i} \right|^2 \tag{8}$$

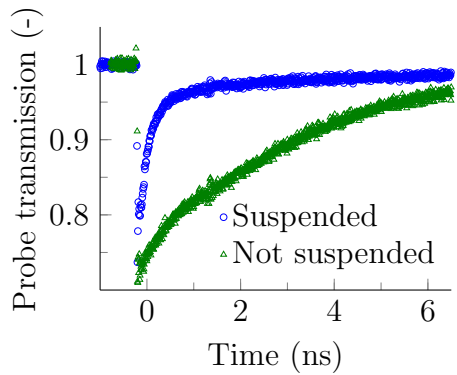

**Figure 7. Drop in free-carrier lifetime.** We measured an increase in the free-carrier recombination rate after the suspension of the silicon beams. Both this finding and the higher propagation losses likely originate in a deterioration of the silicon wires' surface state during the fabrication of the suspended beams.

with $r = f_s \frac{\check{\mathcal{G}}}{4\overline{\gamma}_{\text{tot}}}$ and $\overline{\gamma}_{\text{tot}} = |\overline{\gamma}_{\text{K}} - i\overline{\gamma}_{\text{TPA}} + \overline{\gamma}_{\text{FC}}\overline{P}|$ the *total* nonlinearity, including two-photon absorption and free-carrier effects, $\phi = -\angle\left(\overline{\gamma}_{\text{K}} - i\overline{\gamma}_{\text{TPA}} + \overline{\gamma}_{\text{FC}}\overline{P}\right)$ and $\overline{P}$ the average power in the waveguide. The free-carrier nonlinearity $\overline{\gamma}_{\text{FC}}$ is complex as free carriers modulate both the index and the absorption – both effects create an imprinted sideband on the probe; in addition, $\overline{\gamma}_{\text{FC}}(\Omega)$ depends on the modulation frequency since free carriers do not respond instantaneously. This is a slow dependency, so we take $\overline{\gamma}_{\text{FC}}$ constant in the range of our sweep [14]. This can be shown by solving for the carrier dynamics [45]

$$\partial_t N_c = \frac{\beta_{\text{TPA}}}{2\hbar\omega}P^2 - \kappa_c N_c \qquad (9)$$

in frequency-domain and using the proportionality $\Delta n \propto -N_c$ and $\Delta\alpha \propto N_c$ between both the index and absorption and the carrier concentration [45]. Here we denote $\beta_{\text{TPA}}$ the two-photon absorption coefficient and $\kappa_c$ the free-carrier recombination rate.

Notably, $\phi > 0$ in absence of free carriers ($\overline{\gamma}_{\text{FC}} = 0$). The observed $\phi < 0$ (fig.6a) is thus linked to the presence of free carriers; in addition, we still use $\overline{\gamma}_{\text{tot}} \approx \overline{\gamma}_{\text{K}} \approx 610\,\text{W}^{-1}\text{m}^{-1}$ on the assumption that the Kerr effect remains the dominant background nonlinearity. This is consistent with the observations that (1) in most cases $\mathcal{F}_{\text{max}}\,[\text{dB}] \approx -\mathcal{F}_{\text{min}}\,[\text{dB}]$ and thus $\phi \approx 0$, (2) the background is flat (fig.5&6a) and (3) the Brillouin efficiencies deduced from the XPM experiment are in reasonable agreement with those inferred from the gain experiment (fig.5b).

### Drop in free-carrier lifetime

Using the set-up presented in [16], we measured a significant drop in the free-carrier lifetime in the suspended beams with respect to the regular non-suspended waveguide. Both this finding and the higher propagation losses likely originate in a deterioration of the silicon wires' surface state during the fabrication of the suspended beams.

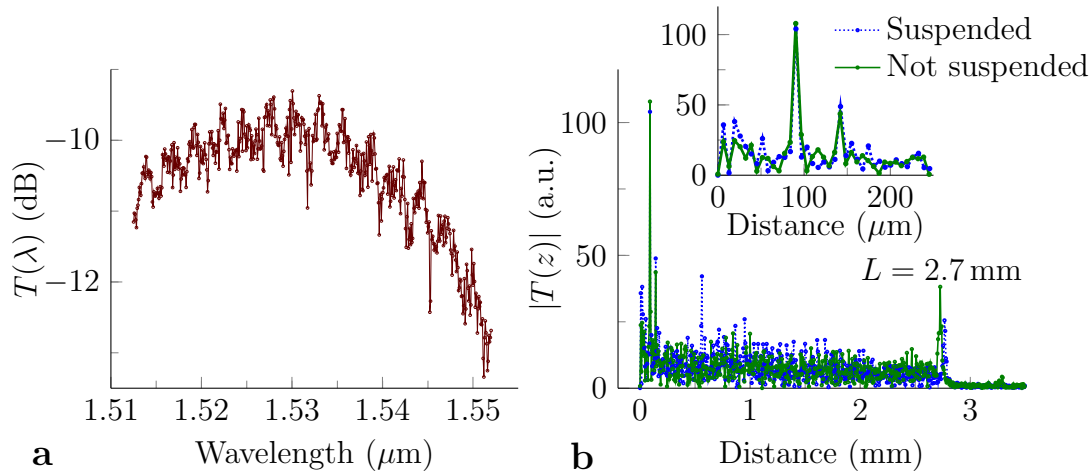

**Figure 8. The interface reflections are small. a**, The fiber-to-fiber transmission spectrum shows the grating bandwidth. It exhibits a slight (1 dB) variation at a free spectral range of about 2.9 nm, both in standard and suspended silicon-on-insulator waveguides. **b**, The Fourier transform of this spectrum reveals potential cavities present in the wire. We compare the regular wire (green) to a wire with 85 suspensions ($L_s = 25.4\,\mu\text{m}$) and anchors ($L_a = 4.6\,\mu\text{m}$) (blue). The distance-axis was calibrated with a group index of $n_g = 4.7$ and shows the waveguide length of 2.7 mm because of grating coupler reflections. These are weaker in the suspended case owing to the higher propagation loss. The 2.9 nm free spectral range (**a**) shows up at $90\,\mu\text{m}$ (**b**, inset). Generally, the two Fourier spectra are nearly identical and do not exhibit notable peaks related to the suspension.

*The interface reflections are negligible*

Our device has discontinuities between the suspended nanobeams and the beams fixed at the anchors (fig.1). Since the beams are spaced periodically, optical reflections may build up. However, we simulated an upper bound for the Fresnel reflection at the discontinuity of less than $10^{-4}$ – indicating that reflections are negligibly small. Empirically, there are indeed no notable differences in the transmission spectrum of a regular waveguide versus that of a suspended waveguide (fig.8). Therefore, our device can be treated as a single-pass structure.

**Acknowledgement**

R.V.L. acknowledges the Agency for Innovation by Science and Technology in Flanders (IWT) for a PhD grant and A.B. the ITN-network cQOM for a postdoc grant. This work was partially funded under the FP7-ERC-InSpectra programme.